\documentclass[usegraphicx]{mn2e}

\def\ltsima{$\; \buildrel < \over \sim \;$}
\def\simlt{\lower.5ex\hbox{\ltsima}}
\def\gtsima{$\; \buildrel > \over \sim \;$}
\def\simgt{\lower.5ex\hbox{\gtsima}}
\def\cgs{{erg cm$^{-2}$ s$^{-1}$}}
\def\ergs{{erg s$^{-1}$}}
\def\cm2{{cm$^{-2}$}}
\def\afe{{photons cm$^{-2}$ s$^{-1}$}}

\def\xrd{{$\chi^{2}_{\rm \nu}$(dof)}}

\def\lum{{$L_{2-10}$}}
\def\lbat{{$L_{14-195}$}}

\def\p1{{Paper I}}

\def\xmm{{\em XMM--Newton}}
\def\chandra~{{\em Chandra}}
\def\beppo{{\em BeppoSAX}}

\def\chandra{{\em Chandra}}

\def\xmm{{\em XMM--Newton}}
\def\szk{{\em Suzaku}}

\def\nh{{N$_{\rm H}$}}

\def\f14{{10$^{-14}$}}
\def\f13{{10$^{-13}$}}
\def\f12{{10$^{-12}$}}
\def\f11{{10$^{-11}$}}
\def\e22{{10$^{22}$}}

\def\feka{{Fe K$\alpha$}}

\def\mrk{{Mrk 231}}

\def\lo{{$L_{\rm [OIV]}$}}
 
\title[Suzaku observation of Mrk 231]{Suzaku reveals X-ray continuum piercing the nuclear absorber in Markarian 231}
\author[Piconcelli et al.]{E.~Piconcelli,$^{1}$\thanks{E-mail: epiconcelli@sciops.esa.int}
G.~Miniutti,$^{2}$
P.~Ranalli,$^{3,4,5}$
C.~Feruglio,$^{6}$
F.~Fiore,$^{7}$
R,~Maiolino$^{8}$
\\
$^{1}$ XMM-Newton Science Operations Centre, ESA, P.O. Box 78, E-28691 Villanueva de la Ca\~nada, Madrid, Spain\\
$^{2}$ Centro de Astrobiologia (CSIC-INTA), Dep. de Astrof\'{i}sica, ESAC, PO Box 78, E-28691, Villanueva de la Ca\~nada, Madrid, Spain\\
$^{3}$ National Observatory of Athens, Palaia Penteli, 15263 Athens, Greece\\
$^{4}$ Osservatorio Astronomico di Bologna (INAF), via Ranzani 1, I--40127, Bologna, Italy\\
$^{5}$ Universit\'a di Bologna, Dipartimento di Astronomia, via Ranzani 1, I-40127, Bologna, Italy\\
$^{6}$ Institute de Radioastronomie Millimetrique (IRAM), 300 Rue de la Piscine,
38406, St. Martin d'Heres, Grenoble, France\\
$^{7}$ Osservatorio Astronomico di Roma (INAF), via Frascati 33, I-00040, Monteporzio Catone, Italy\\
$^{8}$ Cavendish Laboratory, University of Cambridge, 19 J. J. Thomson Avenue, Cambridge, CB3 0HE, UK
}

\begin{document}

\date{Accepted . Received}

\pagerange{\pageref{firstpage}--\pageref{lastpage}} \pubyear{2002}

\maketitle

\label{firstpage}

\begin{abstract}
We report the results from  a 2011 {\it Suzaku} observation of the nearby low-ionization BAL quasar/ULIRG Markarian 231. These data reveal that the X-ray  spectrum has undergone a  large variation from the 2001 \xmm\ and \beppo\ observations. 
We interpret this finding according to a scenario whereby the X-ray continuum source is obscured by a two-component partial-covering absorber with \nh\ $\sim$10$^{22}$ and $\sim$10$^{24}$ \cm2, respectively. The observed spectral change is  mostly explained by a progressive appearance of the primary continuum at $<$ 10 keV due to the decrease of the covering fraction of the denser absorption component.
The properties of the X-ray obscuration in Mrk 231 match well with those of the X-ray shielding gas predicted by the theoretical models for an efficient radiatively-driven acceleration of the BAL wind. In particular, the X-ray absorber might be located at the extreme
base of the outflow.
We  measure a 2-10 keV luminosity of \lum\ = 3.3  $\times$ 10$^{43}$ erg s$^{-1}$ for the 2011 data set, i.e. an increase of 30\% with respect to
the 2001 value.

\end{abstract}


\section{Introduction}
According to  most of the models of merger-driven quasar/massive galaxy co-evolution,
quasars should undergo an heavily obscured phase associated with the presence of massive outflows capable of 
sweeping the bulk of the gas reservoir away from the surrounding host galaxy (e.g. Silk \& Rees 1998; Fabian 1999; Cattaneo et al. 2009; Zubovas \& King 2012).
\mrk\ is regarded as one of the most promising candidates of a quasar that gradually emerges from the obscured phase 
prior to becoming X-ray unobscured/optically bright, while the AGN feedback onto the host galaxy ISM is active.
\mrk\ is indeed unique for many reasons (e.g. Lipari et al. 2009). Firstly, it is the nearest quasar ($z$ = 0.042) and  the most luminous  ($L_{IR}$$\equiv$$L_{8-1000 \mu m}$ = 3.6 $\times$ 10$^{12}$ L$_{\odot}$) ULIRG in the local Universe, with a significant fraction ($>$40\%) of its bolometric luminosity due to starburst activity (SFR $\sim$ 200 M$_\odot$ yr$^{-1}$), with typical signatures of a late-state merger.
Furthermore, \mrk\ belongs to the rare subclass of low-ionization broad absorption line (LoBAL) quasars, which exhibit low-ionization (i.e. MgII and AlIII) BALs and very weak [OIII] emission in their optical spectra. Their unusual properties were explained by Boroson \& Meyers (1992) in terms of absorption  with higher column density  and larger covering fraction than in other BAL quasars, which blocks the ionizing radiation from the nucleus that normally illuminates the narrow-line region.
Supergiant expanding bubbles of ionized gas on kpc-scale have been also detected (see Lipari et al. 2009).

The remarkable discovery of spatially-extended, broad  wings in the CO(J=1-0) emission line profile with interferometric millimeter observations of \mrk\ (Feruglio et al. 2010) revealed the presence of a giant molecular outflow with an estimated mass outflow
rate of $\sim$700 M$_\odot$ yr$^{-1}$, which largely exceeds the ongoing SFR in the host galaxy.
This finding provided one of the very first direct and compelling evidence of quasar-driven feedback removing large amounts of molecular gas, i.e. the material of star formation, from the central regions of the galaxy (then confirmed by  Rupke \& Veilleux 2011, Aalto et al. 2012 and Cicone et al. 2012).
All these results support the hypothesis that in \mrk\ both the nucleus and the host galaxy  are undergoing a key phase, still poorly investigated, in the evolution from an ULIRG/gas-rich system to a standard, X-ray- and optically-bright quasar in a passive galaxy.
 
This motivates to shed further light on the properties of the  central engine in \mrk\ via penetrating X-ray observations. 
However, despite its peculiarity and distinctiveness, \mrk\ has not been targeted by an extensive, well-defined program of X-ray observations so far.
The {\it ASCA} data taken in 1999 suggested the presence of  multiple reprocessed emission components as possible explanation for the very complex spectrum (Maloney \& Reynolds 2000).
Then, four short (40 ks each) exposures have been performed by \chandra\ (Gallagher et al. 2005),  while \xmm\ observed this source for only $\sim$20 ks in 2001. Braito et al. (2004, B04 hereafter) published these data along with those obtained  by  \beppo\ a few months later, revealing the presence of a 3$\sigma$ hard X-ray ($>$ 15 keV) excess in the Phoswich Detector System (PDS) spectrum, suggesting that the X-ray primary continuum is heavily obscured by a Compton-thick (CT) absorber (i.e. \nh\  \simgt\ 1.6 $\times$ 10$^{24}$ \cm2).
This matches well with X-ray studies of LoBAL quasars that typically find highly absorbed spectra (Morabito et al. 2011; Ballo et al. 2011).
B04 derive a quasar-like intrinsic luminosity of \lum\ $\sim$ 10$^{44}$ \ergs\ for the AGN in \mrk.
However, it is important to stress the large uncertainties affecting these data and, therefore, the inferred spectral parameters above 10 keV.

Here, we present the first 0.7-30 keV observation of \mrk\ obtained with \szk.
The unique broad-band coverage of \szk\ allows the simultaneous study of the complex, absorption dominated-spectrum below 10 keV, and the very hard X-ray featureless continuum of the AGN, free from contamination from absorption. 
The \szk\ data confirm the \beppo/PDS detection of a high-energy excess emission above the extrapolation of the 2-10 keV spectrum.
%======================================================
\begin{figure*}
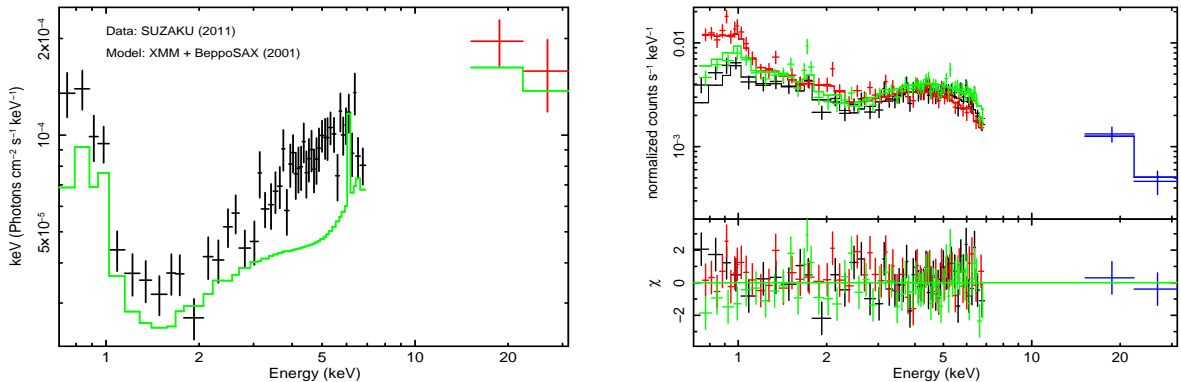

\begin{center}
\includegraphics[width=5cm,height=7.5cm,angle=-90]{epfig1.ps}
\hspace{0.5cm}\includegraphics[width=5cm,height=7.5cm,angle=-90]{epfig2.ps}
\caption{{\bf(a)}--{\it Left:} the unfolded \szk\ XIS0 (black points) $+$ PIN (red points) spectrum of Mrk 231 collected in 2011 compared with the best-fit model to the \xmm\ $+$ \beppo\ data  taken in 2001.
{\bf(b)}--{\it Right:} the  \szk\ spectrum  fitted with the  best-fit 'double partial-covering absorber' model (XIS0: black points; XIS1: red points; XIS3: green points; PIN: blue points). The lower panel shows the deviations of the observed data from the model in units of standard deviation.}
\label{spettri}
\end{center}
\end{figure*}
%==
%======================================================
\section{Observations and data reduction}

\mrk~has been observed  by \szk\ (Mitsuda et al. 2007) on 2011 April 27 (Obs. ID: 706037010) for 200 ks.
During the exposure, X-ray Imaging Spectrometer (XIS) CCDs (e.g. Koyama et al. 2007) and Hard X-ray Detector (HXD; e.g. Takahashi et al. 2007) have been operated in their normal mode, with both front- and back-illuminated XIS chips in the 3$\times$3 and 5$\times$5 editing modes.
Owing to an instrumental failure, XIS2 is unavailable for observations performed after November  2006.
To maximize the HXD effective area, the target has been observed at the HXD nominal aim point.
Data from XIS and HXD have been reduced with  FTOOLS 6.12 and \szk\ software Version 19, using the latest calibration files available (2012 February 11 release) and following the prescriptions described in the \szk\ Data Reduction Guide\footnote{Available at http://suzaku.gsfc.nasa.gov/docs/suzaku/analysis/abc}.

For the XIS detectors (i.e. XIS0, 1 and 3) source counts are extracted from a circular source-free region with a radius of 146
arcsec centered on the peak of the X-ray emission. Background counts are estimated from a similar circular region located close to the source.
The XIS response  and ancillary response  files have been produced using the FTOOLS tasks {\it xisrmfgen}  and {\it xissimarfgen}, respectively.
The  3$\times$3 and 5$\times$5 spectra and matrices have then been combined together using {\it addascaspec}.
All spectra are binned to a minimum of 30 counts per bin to allow use of the $\chi^2$ statistic.
During the \szk~observations \mrk\ did not show any variation in X-ray flux and the spectral shape. 

The HXD/PIN cleaned event file and the corresponding "tuned" non-X-ray background (NXB) event file (ae706037010\_hxd\_pinbgd.evt) 
have been used as input to the  {\it hxdpinxbpi} task. 
The cosmic background spectrum is created with a model provided by the \szk\ team based on the formula by Gruber
et al. (1999) using a flat response (ae\_hxd\_pinflate10\_20101013.rsp) and then combined with the NXB spectrum extracted from  
the NXB event file. 
The resulting combination is used for the background subtraction.
We consider only data in the 15-30 keV energy range, where the source contribution is $\sim$6\% of the total signal of the PIN.
The source contribution drops down to 3\% in the  30-50 keV energy bin.
This choice ensures that the source signal is well above the current systematic uncertainty on the PIN background, i.e.  $\sim$~3\% for observation with exposure longer than 10 ks (Fukazawa et al. 2009).
We use the appropriate response file (ae\_hxd\_pinhxnome10\_20101013.rsp) for HXD-nominal pointing for the spectral analysis.
Here, we present data from PIN only, since no significant signal was detected from GSO.
Relative normalization among XIS0--3 and PIN are set as reported in Maeda et al. (2008) for an observation performed at the HXD nominal position.

\xmm\ and \beppo\ observations of \mrk\ performed in 2001 are also used here.
\xmm\ data have been reduced with SAS 11.0 (Gabriel et al. 2004). The pn source and background spectrum and the matrices are created using standard procedures (e.g., Piconcelli et al. 2004).
The \beppo/PDS spectrum and the corresponding response matrix have been downloaded from the \beppo\ public archive.

Spectra are analyzed with the XSPEC v12.7 package (Arnaud 1996). All fitting models presented here include absorption caused by the line-of-sight Galactic column density of \nh\ = 1.3 $\times$ 10$^{20}$ \cm2\ (Dickey \& Lockman 1990). 

\section{Spectral variability and interpretation}
\label{xmm2001}
 
B04 found a best-fit model (indicated as model  A in their paper) to the joint 2001 \xmm\ and  \beppo\ spectrum of \mrk\ consisting of a combination of a power law modified by a fully-covering CT absorber, and a mix of reprocessed AGN emission absorbed by an additional screen with \nh\ $\approx$ 10$^{22}$ \cm2, and multi-temperature thermal emission due to the starburst activity, which accounts for the very complex 0.5-10 keV spectral shape.

Fig. 1a shows the application of this fitting model  to the \szk\ data taken in 2011. 
This comparison implies a very significant X-ray spectral variation in a 10 yr time span, accompanied with a moderate increase of the flux over the \szk\ bandpass. In particular, the XIS spectral data in the 2-10 keV band  show the largest differences with respect to the \xmm\ ones. Moreover, Fig. 1a  indicates that the normalization of the \feka\ emission line at $\sim$6 keV remains constant between the two epochs (see Sect. 4), suggesting no variation in both spectral features, i.e. fluorescence iron line and continuum, associated with the reflection component (George \& Fabian 1991; Ghisellini et al. 1994), which is believed to originate from a distant reprocessor (typically the visible inner wall  of the obscuring torus itself).
 The soft X-ray band in model A is entirely due to starburst emission. As such, no flux variability has to be expected in the soft X-ray band. The 2-10 keV band is instead dominated by X-ray reflection likely on \simgt\ 0.1 pc scales. Variability in response to continuum changes could in principle be observed on a 10 yr timescale, but this is at odds with the apparent constancy of the associated Fe line intensity. Hence, the only plausible explanation for the soft and hard X-ray band flux variability is that part of the X-ray continuum is visible below 10 keV. 

We then replace the CT absorber of model A with a partial covering one which allows some continuum leakage below 10 keV.
According to this scenario, the primary continuum power law is seen through two distinct, nested  layers of absorption.
The underlying idea is that the observed variation in the \szk\ spectrum with respect to the \xmm\ one can be explained in terms of changes in the physical and geometrical properties of these absorbers. 
In our analysis, we use  the {\tt zxipcf} model in XSPEC (Reeves et al. 2008), which describes partial covering absorption by partially ionized material at the redshift of the source.
This allows us to explore a large variety of possible ionization states and covering fraction of the obscuring medium.
To account for the reflection component, we use the {\tt pexmon} model (Nandra et al. 2007) which includes both continuum reflection (Magdziarz \& Zdziarski 1995) and the associated Fe K line emission (George \& Fabian 1991), with the metal abundances of the reflector fixed to the solar value and its inclination angle fixed to 60$^\circ$. The photon index of the X-ray continuum and reflection is fixed to $\Gamma$ = 1.9, i.e. the typical value measured for quasars (Piconcelli et al. 2005).
Similarly to B04, we also include in the fit two {\tt mekal} thermal components with solar abundance.
This model yields an excellent description of the 2001 \xmm $+$\beppo\ data, i.e. \xrd\ = 0.96(76) (Table \ref{tab:fit}).
The spectral complexity and the quality of the data above 10 keV do not allow to put tight constraints on  
 both the column density and the ionization parameter $\xi$ (expressed in erg cm s$^{-1}$; see Reeves et al. 2008) of the this absorber.
Therefore, we fix the parameters to their best-fit values deduced by the fit, i.e., \nh\ = 1.6 $\times$ 10$^{24}$ \cm2\ and log $\xi$ = -2. The covering fraction of this CT absorber is $\sim$0.85. 
The covering fraction of the second partially-covering absorber is  $\sim$ 0.92 (while fixing $C_f$ to the unity worsens the fit by $\Delta\chi^{2}$ = 26 for 1 dof).
 The best-fit values for the column density and the ionization parameter of this obscuring component are \nh\ = 5.9 $\times$ 10$^{22}$ \cm2 and log $\xi$ = -0.4, respectively. 

In this fit, we assume that the reflection component is not obscured by the two-component absorber and  its normalization $A_R$ is equal to that of the primary continuum $A_{PL}$.
However, a fit with the reflection component obscured only by the outer/Compton-thin absorber produces similar results. Nonetheless, we prefer a solution in which the reflection component is left outside the two-component partial-covering absorber as the constant Fe K line flux suggests that the reflection originates far away the continuum source, while the partial-covering and warm nature of the Compton-thin absorption is more consistent with an inner, nuclear location rather than a pc-scale one.
The value of the reflection fraction is $R$ $\sim$ 0.2 (defined as $\Omega/2\pi$, where $\Omega$ is the solid angle subtended by the reflector for isotropic incident emission).
The temperature of the  two {\tt mekal} components accounting for the thermal diffuse emission is $k$T = 0.4$\pm$0.1  and $k$T = 1.0$\pm$0.2 keV, respectively, i.e. consistent with the values derived by spatially-resolved \chandra\ spectroscopy (Gallagher et al. 2002). 
Finally, this model implies a 2-10(15-30) keV observed luminosity of 2.5(11.9) $\times$  10$^{42}$ \ergs.

%%%%%%%%%%%%%%%%%%%%%%%%%%%%%%%%%%%%%%%% flux 15. 60. kev = 6.6E-12 $^{+}_{-}$
\begin{table*}
\centering
\caption{Best-fitting spectral parameters for  broad-band X-ray spectra of \mrk\ measured assuming the DPCA model (Sect. \ref{xmm2001} and \ref{szk2011}).} 
\label{tab:fit}
%\begin{center}
\begin{tabular}{ccccccccc}
\hline\hline
\multicolumn{1}{c}{Obs./date}&
\multicolumn{1}{c}{$A_{PL}$}&
\multicolumn{1}{c}{\nh/log $\xi$}&
\multicolumn{1}{c}{$C_f$}&
\multicolumn{1}{c} {\nh/log$\xi$}&
\multicolumn{1}{c}{$C_f$}&
\multicolumn{1}{c}{$R$}&
\multicolumn{1}{c}{Flux}&
\multicolumn{1}{c}{$\chi^{2}_{\rm \nu}$(dof)}
\\
 (1)&(2)&(3)&(4)&(5)&(6)&(7)&(8)&(9)\\\hline\\ 
{\it XMM}(6/2001) $+$  & 2$\pm$1&160$^\dag$/-2$^\dag$&0.85$^{+0.05}_{-0.14}$& 5.9$^{+1.2}_{-2.0}$/-0.4$^{+0.7}_{-0.8}$&0.92$\pm$0.02&0.19$^{+0.24}_{-0.12}$&1.1/6.4/30.2&0.96(76)\\
{\it SAX}(12/2001)& &&&&&&&\\ 
& &&&&&&&\\ 
\szk(4/2011) & 2.9$\pm$0.8&160$^\dag$/-2$^\dag$&0.78$^{+0.05}_{-0.08}$&6.3$^{+0.5}_{-0.6}$/-0.5$^{+0.2}_{-0.4}$ &0.95$\pm$0.01&0.08$^\dag$&1.7/10.1/37.1&1.09(285)\\ 
\hline
\end{tabular}
%\end{center}
\begin{quote} 
%The columns give the following information:
Notes: (1) X-ray telescope (epoch);
(2) Intensity of the power-law primary continuum (in units of 10$^{-3}$ photons keV$^{-1}$ cm$^{-2}$ s$^{-1}$);
(3) Column density (in units of 10$^{22}$ \cm2) and ionization parameter of the CT absorber;
(4) Covering fraction of the CT absorber 
(5) Column density (in units of 10$^{22}$ \cm2) and ionization parameter  of the Compton-thin absorber;
(6) Covering fraction of the  Compton-thin absorber;
(7) Reflection fraction; 
(8) Flux in the 0.5--2/2--10/15--30 keV band (in units of 10$^{-13}$ \cgs). 
 $^\dag$ The parameter has been fixed to the best-fitting value deduced by the fit.
%\end{center}
\end{quote} 
\end{table*}
%%%%%%%%%%%%%%%%%%%%%%%%%%%%%%%%%%%%%%%%%%%%% $^{+ }_{- }$
%%%%%%%%%%%%%%%%%%%%%%%%% $^{+ }_{- }$
\section{Broad-band Suzaku spectroscopy}
\label{szk2011}

As the next step, we apply the double partial-covering absorber (DPCA hereafter) model that has been found to successfully fit the 2001 \xmm $+$\beppo\ data to the broad-band \szk\ spectrum taken in 2011. 
Our goal is to test if both data sets can be described by the same model, and the observed variability can be ascribed to variations in the line-of-sight absorbers. 
The normalization of \feka\ emission line measured from the \szk\ spectrum is   $A_{Fe}$ = 9$\pm$6 $\times$ 10$^{-7}$ \afe.
Such a value is fully consistent with the \xmm\ one ($A_{Fe}$ = 1.5$\pm$1.1 $\times$ 10$^{-6}$ \afe) and those inferred from \chandra\ observations (e.g. Gallagher et al. 2005).
% i.e. $\langle$$A_{Fe}$$\rangle$ = 0.9$^{+0.7}_{-0.8}$ \afe.
The \feka\ line and, hence, the associated reflection continuum therefore remains approximately unchanged over these epochs (see Fig. 1a).
 Accordingly, the value of reflection fraction $R$ has been forced to vary within the 90\% confidence level interval derived for these parameter from the 2001 data, keeping the same normalization ($A_R$ = 2 $\times$ 10$^{-3}$ photons keV$^{-1}$ cm$^{-2}$ s$^{-1}$). We then fix $R$ =  0.08, i.e., the best-fit value deduced by the fit.
The values of the \nh\ and $\xi$ of the CT absorber are constrained as for the \xmm $+$\beppo\ spectrum. 
The result of the DPCA fit to the 0.7--30 keV \szk\ data is shown in Fig. 1b. This model provides a very good fit with \xrd\ = 1.09(285).

As expected on the basis of Fig.~\ref{spettri}a, we measure an increase of the flux in the 0.5-2, 2-10 and 15-30 keV bands (e.g. Table \ref{tab:fit}) from 2001 to 2011. It mostly results from the combination of a larger normalization of the power law  of the primary continuum and a smaller covering fraction of the CT absorber in the \szk\ spectrum (although, given their large uncertainties, both parameters are however consistent at 90\% with those inferred in Sect. \ref{xmm2001}).
The best-fitting temperatures  ($k$T = 0.23$^{+0.18}_{-0.06}$ and $k$T = 1.26$^{+0.16}_{-0.22}$ keV) and the normalizations of  the two {\tt mekal} soft X-ray emission components are in good agreement with those derived from the \xmm\ spectrum, as expected for spectral component associated with starburst emission and/or photoionized gas emission.
Once model DPCA is assumed, we estimate a 2-10(0.5-30) keV flux of $\sim$1(5.7) $\times$ 10$^{-12}$ \cgs, corresponding to an observed luminosity of $\sim$4(22) $\times$ 10$^{42}$ \ergs.

Finally, replacing the warm absorbers in the DPCA model with  cold ones yields a worse fit to the \szk\ data, with an associated \xrd\ = 1.28(285), and large positive residuals around 5 keV where it predicts less spectral curvature and a continuum less intense than observed.

\section{Discussion}

The analysis of the \szk\ data presented here has revealed a large variation of the X-ray spectral shape from the last X-ray observations of \mrk\ performed with \xmm\ and \beppo\ in 2001 (Fig. \ref{spettri}a).
We have successfully fitted the broad-band \szk\ spectrum with a DPCA model. This model also provides an excellent fit to the 2001 data. 
In this scenario,  two absorbers lie along our line of sight (LoS) to the nucleus of \mrk: one 
with a CT column density, and another one  with  \nh\ $\sim$ a few of  10$^{22}$ \cm2.
Both absorbers are slightly ionized and have a covering fraction less than 1. This allows to detect the leakage below 10 keV of the direct emission from the X-ray source. 
Accordingly,  the spectral variation shown in Fig. \ref{spettri}a can be basically explained by a progressive appearance of the primary continuum at energies $<$10 keV due to a decrease of the covering fraction of the CT absorber, suggesting that this absorption component is the innermost, more variable one.
This interpretation appears to be consistent both with (i) the spectral properties inferred by 
previous X-ray observations, as well as (ii) the LoBAL nature of the quasar in \mrk.

Maloney \& Reynolds (2000) indeed proposed the existence of two layers of obscuration after the analysis of a moderately long {\it ASCA} exposure.  However, they assumed that no fraction of the X-ray primary continuum is visible in the  {\it ASCA} range, i.e. the quasar central engine is completely buried.
 If, as we suggest, the $C_f$ variation occurred between 2001 and 2011, it is indeed possible that {\it ASCA} caught the source at an epoch when the $C_f$ of the CT absorber was close to unity.
However, it  is important to bear in mind that the limited bandpass of {\it ASCA} did not allow to put any meaningful constraint on the CT obscured X-ray continuum.
 The variability in the hard X-ray flux detected during a 40 ks \chandra\ observation led Gallagher et al. (2002) to question the reflection-dominated hypothesis for the X-ray spectrum of this quasar, although 
such variability was not detected in any of the subsequent three similar  \chandra\ exposures performed in 2003.
They suggested the existence of a third spectral component, i.e. an absorbed   (\nh\ $\sim$ 3 $\times$ 10$^{22}$ \cm2) power law,  in addition to the thermal plasma and the reflection continuum, which may arise from a compact X-ray scattering medium.
Gallagher et al. (2005) found an acceptable fit to the combined 0.5-8 keV spectrum from the four \chandra\ observations with a model consisting of a power law modified by three absorbers plus soft thermal emission, which may resemble our DPCA model in terms of overall spectral shape in this energy band, 'weak'  \feka\ emission  and, mostly, for the  presence of a multi-component absorber.
Furthermore, the consistency of the flux of the \feka\ emission line over  2001-2003 found by \chandra\ and \xmm\  observations of \mrk\
 is suggestive of a Compton reflection component likely to be associated with distant, optically thick material, and, remarkably, the 2011 \szk\ data lend further support to this scenario.
The DPCA model proposed here is therefore able to explain in a coherent fashion the several pieces of evidence collected over the years thanks to the previous  limited bandpass and/or less sensitive X-ray observations. 
The hard X-ray flux values measured from the \chandra\ observations are consistent each other and are slightly larger (i.e. consistent within 2$\sigma$) than the \xmm\ value (Gallagher et al. 2005). In the proposed DPCA scenario, this can be interpreted as lack of large variations in the physical properties of the CT absorber.

The presence of the highly absorbed power-law emission at $E >$ 15 keV above the extrapolation of the 2-10 keV spectrum claimed by B04  thanks to the 2001 \beppo\ observation is fully confirmed by the present \szk\ data. X-ray spectral properties of LoBAL 
are very scarce due to their weakness in the 0.5-10 keV range (Green et al. 2001; Morabito et al. 2011), interpreted in terms of
heavy obscuration. Even much more extraordinary is the detection of these objects at higher energies: as far we know,
\mrk\ is the only member of this class of AGN for which a spectrum has been collected in the band above 10 keV.

The change in the spectral shape  discovered here as well as the short-timescale variability in the
hard-band emission reported by Gallagher et al. (2002) (assuming an origin in the multi-component absorber) seems to indicate that the  bulk of the X-ray obscuration likely occurs in a compact region much closer to the X-ray source than the canonical 0.1-1 pc-scale  torus invoked in the Unified model, or extended structures in the complex circumnuclear environment of \mrk\ (e.g. Klockner et al. 2003, and references therein).
Both ionized and neutral X-ray absorbers with large column densities are commonly observed in BAL quasars (e.g. Green et al. 2001; Piconcelli et al. 2005; Morabito et al. 2011; Giustini et al. 2011)
They have been indeed associated with the BAL outflow (e.g. Elvis 2000), which is likely launched
from the accretion disk surrounding the supermassive black hole (Young et al. 2007), and their variability explained in terms of outflowing clouds crossing  our LoS.
In particular, it has been proposed that this wind must be shielded from the X-ray continuum source to prevent the BAL
absorbing clouds from becoming too highly ionized. This would inhibit a very efficient mechanism 
of gas acceleration such as the radiation-driven acceleration, by eliminating most of the UV transitions
that absorb the photons necessary to accelerate the outflowing gas (Murray et al. 1995; Proga 2007, and
references therein). This X-ray-shielding region must be effectively opaque to X-rays, as well as sufficiently ionized (and
dust-free) to allow UV line opacity, and might represent the inner section of the wind itself (and/or located at the extreme base of the outflow).
The properties of the DPCA discovered in \mrk\ match well with those predicted for the shielding gas by the BAL outflow models, providing an important observation-based evidence for the theory of physical mechanism responsible for launching and accelerating AGN winds.
These nuclear, highly supersonic outflows are believed to compress the surrounding medium and generate a shock, which moves outwards and sweeps the surrounding medium out, i.e. {\it the AGN  feedback mechanism} (Menci et al. 2008; Zubovas \& King 2012).

The presence of a partially-covering absorber suggests that the typical size of the X-ray 
absorber along our LoS is comparable with that of the X-ray continuum source.
The latter is likely of the order of $\sim$10 $R_G$, as estimated via occultation or microlensing in a few AGNs
 (Dai et al. 2010; Risaliti et al. 2007; Maiolino et al. 2010).
Accordingly, the DPCA is not the torus envisaged by the Unification model, which can be however present in \mrk, as it should lie  (i) on larger scales and (ii) edge-on with respect to our LoS. 
The non-variable reflection emission component observed in the X-ray spectrum of \mrk\ is likely associated to this obscuring/reprocessing medium located distant from the central engine.

Finally, the broad bandpass of \szk\ allows us to measure the intrinsic luminosity of the X-ray source in \mrk. We find an intrinsic  luminosity of \lum\ = 3.3 $\times$  10$^{43}$ \ergs\ in the 2-10 keV energy range, while the luminosity in the 15-30 keV PIN band is 1.7  $\times$  10$^{43}$ \ergs.
Using the DPCA model for the joint \xmm\ $+$ \beppo\ spectrum, we calculate a \lum\ = 2.6 $\times$  10$^{43}$  \ergs. 
It is interesting to compare this X-ray luminosity with that expected on the basis of well-established isotropic indicator of AGN luminosity.
The luminosity of the mid-IR emission line [O IV] 25.89$\mu$m  has been shown to be one of the most accurate proxies of \lum\ (Melendez et al. 2008).  
Armus et al. (2007) reported an upper limit for the  [O IV] 25.89$\mu$m luminosity of \lo\ $<$ 2.64 $\times$  10$^{41}$ \ergs\ based on 2004 {\it Spitzer} data.
Using the  \lo--\lum\ relation in Melendez et al. (see Table 3), the predicted 2-10 keV luminosity for \mrk\  is $<$ 3.7 $\times$  10$^{43}$ \ergs.
This matches well with the \lum\ values deduced from both \szk\ and \xmm\ observations.
Similar results are also obtained by the comparison of the luminosity  in the 14-195 keV range predicted on the basis of \lo, i.e. \lbat\ $<$ 10$^{44}$ \ergs,
and those estimated from \szk\ and \xmm\ $+$ \beppo\ data, i.e. \lbat\ = 7 and 5.7 $\times$  10$^{43}$ \ergs, respectively.

Future X-ray observations with broad-band X-ray capabilities (i.e., {\it NuSTAR} and {\it ASTRO-H}) as well as well-defined monitoring program with present-day X-ray telescopes
will allow to provide much more detailed insights  into the kinetic, geometrical and physical properties of the nuclear absorber(s) in this composite ULIRG/LoBAL quasar, where AGN-driven outflows in the neutral, ionized and molecular gas
capable of quenching star formation in the host galaxy have been found.

%======================================================

\section*{Acknowledgments}
We thank the referee for useful comments and suggestions that helped improve the quality and the presentation of the paper.
We are grateful to all the {\it Suzaku} team members.
EP would like to acknowledge support by INSA. GM thanks the Spanish Ministry of Science and Innovation for partial support through grant
AYA2010-21490-C02-02. PR acknowledges financial contribution from the agreement ASI-INAF
I/009/10/0, and a grant from the Greek General Secretariat of Research
and Technology in the framework of the program Support of Postdoctoral
Researchers. Based on observations obtained with {\it XMM-Newton}, an ESA science mission with instruments and contributions directly funded by ESA Member States and NASA.


\begin{thebibliography}{99}

\bibitem[\protect\citeauthoryear{Aalto et 
al.}{2012}]{2012A&A...537A..44A} Aalto S., Garcia-Burillo S., Muller S., Winters J.~M., van der Werf P., Henkel C., Costagliola F., Neri R., 2012, A\&A, 537, A44 


\bibitem[\protect\citeauthoryear{Armus et al.}{2007}]{2007ApJ...656..148A} 
Armus L. et al., 2007, ApJ, 656, 148 


\bibitem[\protect\citeauthoryear{Arnaud}{1996}]{1996ASPC..101...17A} Arnaud 
K.~A., 1996, ASPC, 101, 17 


\bibitem[\protect\citeauthoryear{Ballo et al.}{2011}]{2011MNRAS.415.2600B} 
Ballo L., Piconcelli E., Vignali C., Schartel N., 2011, MNRAS, 415, 2600 


\bibitem[\protect\citeauthoryear{Boroson 
\& Meyers}{1992}]{1992ApJ...397..442B} Boroson T.~A., Meyers K.~A., 1992, ApJ, 397, 442 


\bibitem[\protect\citeauthoryear{Braito et 
al.}{2004}]{2004A&A...420...79B} Braito V. et al., 2004, A\&A, 420, 79 (B04)


\bibitem[\protect\citeauthoryear{Cattaneo et 
al.}{2009}]{2009Natur.460..213C} Cattaneo A. et al., 2009, Nat, 460, 213 


\bibitem[\protect\citeauthoryear{Cicone et 
al.}{2012}]{2012A&A...543A..99C} Cicone C., Feruglio C., Maiolino R., Fiore F., Piconcelli E., Menci N., Aussel H., Sturm E., 2012, A\&A, 543, A99 


\bibitem[\protect\citeauthoryear{Dai et al.}{2010}]{2010ApJ...709..278D} 
Dai X., Kochanek C.~S., Chartas G., Koz{\l}owski S., Morgan C.~W., Garmire 
G., Agol E., 2010, ApJ, 709, 278 


\bibitem[\protect\citeauthoryear{Dickey 
\& Lockman}{1990}]{1990ARA&A..28..215D} Dickey J.~M., Lockman F.~J., 1990, ARA\&A, 28, 215 


\bibitem[\protect\citeauthoryear{Elvis}{2000}]{2000ApJ...545...63E} Elvis 
M., 2000, ApJ, 545, 63 


\bibitem[\protect\citeauthoryear{Fabian}{1999}]{1999MNRAS.308L..39F} Fabian 
A.~C., 1999, MNRAS, 308, L39 


\bibitem[\protect\citeauthoryear{Feruglio et 
al.}{2010}]{2010A&A...518L.155F} Feruglio C., Maiolino R., Piconcelli E., Menci N., Aussel H., Lamastra A., Fiore F., 2010, A\&A, 518, L155 


\bibitem[\protect\citeauthoryear{Fukazawa et 
al.}{2009}]{2009PASJ...61S..17F} Fukazawa Y. et al., 2009, PASJ, 61, 17 


\bibitem[\protect\citeauthoryear{Gabriel et 
al.}{2004}]{2004ASPC..314..759G} Gabriel C. et al., 2004, ASPC, 314, 759 


\bibitem[\protect\citeauthoryear{Gallagher et 
al.}{2002}]{2002ApJ...569..655G} Gallagher S.~C., Brandt W.~N., Chartas G., 
Garmire G.~P., Sambruna R.~M., 2002, ApJ, 569, 655 


\bibitem[\protect\citeauthoryear{Gallagher et 
al.}{2005}]{2005ApJ...633...71G} Gallagher S.~C., Schmidt G.~D., Smith 
P.~S., Brandt W.~N., Chartas G., Hylton S., Hines D.~C., Brotherton M.~S., 
2005, ApJ, 633, 71 


\bibitem[\protect\citeauthoryear{George 
\& Fabian}{1991}]{1991MNRAS.249..352G} George I.~M., Fabian A.~C., 1991, MNRAS, 249, 352 


\bibitem[\protect\citeauthoryear{Ghisellini, Haardt, 
\& Matt}{1994}]{1994MNRAS.267..743G} Ghisellini G., Haardt F., Matt G., 1994, MNRAS, 267, 743 


\bibitem[\protect\citeauthoryear{Giustini et 
al.}{2011}]{2011A&A...536A..49G} Giustini M. et al., 2011, A\&A, 536, A49 


\bibitem[\protect\citeauthoryear{Green et al.}{2001}]{2001ApJ...558..109G} 
Green P.~J., Aldcroft T.~L., Mathur S., Wilkes B.~J., Elvis M., 2001, ApJ, 
558, 109 


\bibitem[\protect\citeauthoryear{Gruber et al.}{1999}]{1999ApJ...520..124G} 
Gruber D.~E., Matteson J.~L., Peterson L.~E., Jung G.~V., 1999, ApJ, 520, 
124 


\bibitem[\protect\citeauthoryear{Kl{\"o}ckner, Baan, 
\& Garrett}{2003}]{2003Natur.421..821K} Kl{\"o}ckner H.-R., Baan W.~A., Garrett M.~A., 2003, Nat, 421, 821 


\bibitem[\protect\citeauthoryear{Koyama et al.}{2007}]{2007PASJ...59S..23K} 
Koyama K. et al., 2007, PASJ, 59, 23 


\bibitem[\protect\citeauthoryear{Lipari et al.}{2009}]{2009MNRAS.392.1295L} 
Lipari S. et al., 2009, MNRAS, 392, 1295 

\bibitem[\protect\citeauthoryear{Maeda et al.}{2008}]{}
 Maeda Y. et al., 2008, Suzaku Memo 2008-06: Recent update of the XRT response: III. Effective Area, available at http://www.astro.isas.jaxa.jp/suzaku/doc/suzakumemo/suzakumemo-2008-06.pdf 

\bibitem[\protect\citeauthoryear{Magdziarz 
\& Zdziarski}{1995}]{1995MNRAS.273..837M} Magdziarz P., Zdziarski A.~A., 1995, MNRAS, 273, 837 


\bibitem[\protect\citeauthoryear{Maiolino et 
al.}{2010}]{2010A&A...517A..47M} Maiolino R. et al., 2010, A\&A, 517, A47 


\bibitem[\protect\citeauthoryear{Maloney 
\& Reynolds}{2000}]{2000ApJ...545L..23M} Maloney P.~R., Reynolds C.~S., 2000, ApJ, 545, L23 


\bibitem[\protect\citeauthoryear{Mel{\'e}ndez et 
al.}{2008}]{2008ApJ...682...94M} Mel{\'e}ndez M. et al., 2008, ApJ, 682, 
94 


\bibitem[\protect\citeauthoryear{Menci et al.}{2008}]{2008ApJ...686..219M} 
Menci N., Fiore F., Puccetti S., Cavaliere A., 2008, ApJ, 686, 219 


\bibitem[\protect\citeauthoryear{Mitsuda et 
al.}{2007}]{2007PASJ...59S...1M} Mitsuda K. et al., 2007, PASJ, 59, 1 


\bibitem[\protect\citeauthoryear{Morabito et 
al.}{2011}]{2011ApJ...737...46M} Morabito L.~K., Dai X., Leighly K.~M., 
Sivakoff G.~R., Shankar F., 2011, ApJ, 737, 46 


\bibitem[\protect\citeauthoryear{Murray et al.}{1995}]{1995ApJ...451..498M} 
Murray N., Chiang J., Grossman S.~A., Voit G.~M., 1995, ApJ, 451, 498 


\bibitem[\protect\citeauthoryear{Nandra et al.}{2007}]{2007MNRAS.382..194N} 
Nandra K., O'Neill P.~M., George I.~M., Reeves J.~N., 2007, MNRAS, 382, 194 


\bibitem[\protect\citeauthoryear{Piconcelli et 
al.}{2005}]{2005A&A...432...15P} Piconcelli E., Jimenez-Bail{\'o}n E., Guainazzi M., Schartel N., Rodr{\'{\i}}guez-Pascual P.~M., Santos-Lle{\'o} M., 2005, A\&A, 432, 15 


\bibitem[\protect\citeauthoryear{Piconcelli et 
al.}{2004}]{2004MNRAS.351..161P} Piconcelli E., Jimenez-Bail{\'o}n E., 
Guainazzi M., Schartel N., Rodr{\'{\i}}guez-Pascual P.~M., Santos-Lle{\'o} 
M., 2004, MNRAS, 351, 161 


\bibitem[\protect\citeauthoryear{Proga}{2007}]{2007ASPC..373..267P} Proga 
D., 2007, ASPC, 373, 267 


\bibitem[\protect\citeauthoryear{Reeves et al.}{2008}]{2008MNRAS.385L.108R} 
Reeves J., Done C., Pounds K., Terashima Y., Hayashida K., Anabuki N., 
Uchino M., Turner M., 2008, MNRAS, 385, L108 


\bibitem[\protect\citeauthoryear{Risaliti et 
al.}{2007}]{2007ApJ...659L.111R} Risaliti G., Elvis M., Fabbiano G., Baldi 
A., Zezas A., Salvati M., 2007, ApJ, 659, L111 


\bibitem[\protect\citeauthoryear{Rupke 
\& Veilleux}{2011}]{2011ApJ...729L..27R} Rupke D.~S.~N., Veilleux S., 2011, ApJ, 729, L27 


\bibitem[\protect\citeauthoryear{Silk 
\& Rees}{1998}]{1998A&A...331L...1S} Silk J., Rees M.~J., 1998, A\&A, 331, L1 


\bibitem[\protect\citeauthoryear{Takahashi et 
al.}{2007}]{2007PASJ...59S..35T} Takahashi T. et al., 2007, PASJ, 59, 35 


\bibitem[\protect\citeauthoryear{Young et al.}{2007}]{2007Natur.450...74Y} 
Young S., Axon D.~J., Robinson A., Hough J.~H., Smith J.~E., 2007, Nat, 
450, 74 


\bibitem[\protect\citeauthoryear{Zubovas 
\& King}{2012}]{2012ApJ...745L..34Z} Zubovas K., King A., 2012, ApJ, 745, L34 

\end{thebibliography}
\end{document}